\documentclass[authoryear,12pt,3p]{jowarticle}

\usepackage[english]{babel}
\usepackage[utf8]{inputenc}
\usepackage{amsmath}
\usepackage{graphicx}
\usepackage{setspace}
\usepackage[colorinlistoftodos]{todonotes}
\usepackage{natbib}

\makeatletter
\renewcommand\section{\@startsection{section}{1}{\z@}{-3.25ex plus -1ex minus -.2ex}{1.5ex plus .2ex}{\normalsize\bf}}
\renewcommand\subsection{\@startsection{subsection}{2}{\z@}{-3.25ex plus -1ex minus -.2ex}{1.5ex plus .2ex}{\normalsize\bf}}
\renewcommand\subsubsection{\@startsection{subsubsection}{3}{\z@}{-3.25ex plus -1ex minus -.2ex}{1.5ex plus .2ex}{\normalsize\bf}}

\usepackage{enumerate}
\usepackage{graphicx}
\usepackage[round]{natbib}
\usepackage{amssymb,amsmath, amscd, amsthm, mathrsfs}

\usepackage{ bbold }
\theoremstyle{definition}

\theoremstyle{remark}

\numberwithin{equation}{section}

\makeatother

\begin{document}
\begin{frontmatter}

\title{Mathematical Responses to the Hole Argument: Then and Now}
\author{Clara Bradley}\ead{cbradle1@uci.edu}
\author{James Owen Weatherall}\ead{weatherj@uci.edu}
\address{Department of Logic and Philosophy of Science \\ University of California, Irvine}

\date{\today}

\begin{abstract}
We argue that several apparently distinct responses to the hole argument, all invoking formal or mathematical considerations, should be viewed as a unified ``mathematical response''.  We then consider and rebut two prominent critiques of the mathematical response before reflecting on what is ultimately at issue in this literature.  
\end{abstract}
\end{frontmatter}
\doublespacing

\section{Introduction}

In a recent paper, \citet{WeatherallHoleArg} contends that careful attention to mathematical practice reveals that the ``hole argument'', familiar from the foundations of general relativity (GR), is fallacious.\footnote{We briefly state the hole argument in section \ref{sec:holeArg}.}  At issue is whether the hole argument, by implicitly invoking a privileged identification of spacetime points in distinct models of GR, thereby makes use of mathematical structure (something like ``point identity”) that those models should not be understood to possess.  This argument has been controversial.  Some, such as \citet{Fletcher} have contended that yes, the hole argument is blocked by these considerations;\footnote{\citet{CurielHoleArg} and \citet{Halvorson+Manchak} apparently endorse this claim as well; see also \citet{Bradley+Weatherall}.}  whereas others, such as \citet{Arledge+Rynasiewicz}, \citet{Roberts}, and \citet{Pooley+Read} deny it.

Weatherall's response to the hole argument is sometimes characterized as ``mathematical'' (as opposed to ``metaphysical'').  But it is not the only ``mathematical'' response in the literature.  Shortly after the hole argument was introduced to the philosophy of physics literature by \citet{Earman+Norton}, \citet{Mundy} offered an apparently different mathematical response, according to which considerations from formal semantics should be seen to block the hole argument.  And more recently, largely in parallel to the literature responding to Weatherall, \citet{Shulman} has observed that the hole argument cannot be expressed in Homotopy Type Theory with Univalent Foundations (HoTT/UF), a recent program in the foundations of mathematics (\cite{HoTT}).\footnote{\citet{Ladyman+Presnell} offer a helpful elaboration and clarification of Shulman's argument.} All of these arguments agree on at least one claim: prior to any questions about metaphysics, the hole argument should be rejected as mathematically illegitimate.\footnote{Another broadly ``mathematical'' response is due to \citet{Iftime+Stachel}, but to relate that paper to those discussed here would require more space than we have.}   

The goal of the present paper is to explore the relationship between Mundy’s response to the hole argument and the more recent responses by Weatherall and Shulman.  We argue that all three arguments are substantially the same, and that considering them together helps isolate the core consideration behind all of them: the hole argument requires one to express facts about models of GR that cannot be expressed in ``the language of GR'', on \emph{any} reasonable understanding of what that could mean.\footnote{\citet[fn. 4] {WeatherallHoleArg} resists parallels with Mundy, but for spurious reasons.} Instead, one must move outside of GR, by augmenting the theory in some way, or by invoking the resources of a background metatheory, such as set theory. This point can be made by regimenting the ``language of GR'' via a higher-order axiomatization, as Mundy does; or it may be made by appeal to HoTT/UF, as Shulman does; or it may be made by appeal to mathematical practice, as Weatherall does.  Perhaps most importantly, in all three cases, the project is not to solve an interpretational problem by adopting some novel formal apparatus; rather, it is to argue that the (formal) problem allegedly raised by the hole argument is illusory.

The remainder of the paper will proceed as follows. In the next section, we will briefly review the hole argument and comment on ``metaphysical'' responses.  Then, we will introduce the three ``mathematical'' responses under consideration.  We will focus on Mundy's response and use it to interpret Weatherall's and Shulman's.  We will then consider two criticisms of the mathematical response, from \citet{RynasiewiczSyntax} and \citet{Roberts}.  We conclude by discussing what the mathematical response really amounts to, and relating it to other issues connected to reference and representation.

\section{The Hole Argument}\label{sec:holeArg}

The hole argument, at least in its modern form, invokes the following mathematical considerations.\footnote{The ``modern'' hole argument is most often associated with \citet{Earman+Norton}; but it has a prehistory both in Einstein's early work on GR \citep{Stachel,Norton1984} and in more philosophical work by  \citet{SteinPrehistory} and \citet{Earman1977} on Leibniz \citep[see][]{WeatherallStein}.} Given a model of GR, which for these purposes is a pseudo-Riemannian manifold, $(M, g_{ab})$, one can construct another model $(M,g_{ab}')$, isometric to the first, by fixing any diffeomorphism $\psi: M \rightarrow M$ and defining $g_{ab}'$ to be the pushforward of $g_{ab}$ along $\psi$, $\psi_*(g_{ab})$.\footnote{For background on GR and the conventions adopted here, see \citet{MalamentGR}.} If the diffeomorphism is not the identity on $M$, then these models disagree at certain points on the value of the metric; and yet, they are isometric by construction, and therefore they are equivalent as pseudo-Riemannian manifolds.  

These facts are widely taken to present a dilemma.  It appears that $(M,g_{ab})$ and $(M,g_{ab}')$ are distinct models that, by virtue of being isometric, agree on all ``observable'' structure.  Thus, if one takes distinct models of GR to correspond to distinct physical situations, the theory distinguishes situations that have no observable differences.  Worse, since $\psi$ may be the identity outside of some precompact set, it seems to follow that the metric outside of that set cannot ``determine'' the metric within it.  And so, the argument goes, a \textit{manifold substantivalist} -- that is, a certain kind of ``realist'' about GR who accepts that distinct models of the theory, in the sense just described, reflect distinct possibilities -- incurs significant costs.

Most authors responding to the hole argument take the basic issue to be metaphysical.\footnote{For up-to-date reviews of metaphysical reactions of the hole argument, see \citet{PooleyOxford} or \citet{NortonSEP}.  \citet{Roberts+Weatherall} discuss the more recent literature.} They agree that GR presents us with distinct models of the sort contemplated here, and that these naturally represent distinct situations; to avoid the dismal conclusions, a would-be realist (or substantivalist) must adopt a metaphysical view on space and time that denies that these putatively distinct possibilities are genuinely different. Of particular concern has been whether or not one's metaphysics is compatible with the doctrine of \textit{Leibniz Equivalence}, according to which any pair of isometric pseudo-Riemannian manifolds represent the same possible world. Manifold substantivalists reject Leibniz equivalence; many authors have developed flavors of \textit{sophisticated substantivalism}, or qualified realism, according to which one can accept Leibniz Equivalence while still endorsing some form of substantivalism. 

\section{The Mathematical Response}\label{sec:mathematics}

As we noted in the Introduction, there is another tradition of responses to the hole argument that seeks to avoid the metaphysical considerations just discussed. Instead, they argue, one should reject the background mathematical claims that others take to generate the hole argument.  Roughly: one should deny that there is any relevant sense in which the construction used in the hole argument generates ``distinct'' pseudo-Riemannian manifolds.  Thus, even a realist who takes distinct models to correspond to distinct possibilities can block the hole argument.

\cite{Mundy} defends this position by showing that pseudo-Riemannian geometry admits an ``intrinsic axiomatization'' in higher-order logic.  Formal semantics then provides a precise characterization of a model of pseudo-Riemannian geometry as representable by a set-theoretic structure, of a particular signature, satisfying the axioms of the theory under some interpretation; it also determines a notion of isomorphism between such structures. The notion of isomorphism one arrives at from his intrinsic axiomatization -- and, indeed, by \emph{any} reasonable axiomatization -- coincides with isometry on other presentations of the theory.\footnote{Mundy also offers a ground-clearing argument that coordinate-based formulations of pseudo-Riemannian geometry are semantically ambiguous.  We focus here on the second, positive part of his argument.}  

Mundy argues that formal theories of this kind only determine their models up to isomorphism, in the precise sense that any statement in the language of the theory that is true in a structure $S$ is true in any structure $S'$ isomorphic to $S$.  In other words, no sentence in the theory can distinguish $S$ and $S'$.  Thus isomorphic models are ``semantically equivalent'' (p. 517), which is equivalence ``in the strongest possible sense'' (p. 519).  But the models at issue in the hole argument, by virtue of being isometric, are isomorphic in this sense. He concludes that these models do not present putatively distinct possibilities, as required in the hole argument, because they are not distinct models. 

What led others to think that Leibniz equivalence was a substantive doctrine? Mundy offers the following diagnosis.  The semantic equivalence of set-theoretic structures representing models of pseudo-Riemannian geometry means that there are no differences between the structures that can be expressed within the ``language of GR''.\footnote{Or at least, the language of pseudo-Riemannian manifolds.  GR would presumably extend this language, but not in ways that affect the argument.}   But one can also step outside of the theory and, by invoking the further expressive resources of the ``semantic metalanguage'' -- basically, ZFC set theory -- one \emph{can} express differences between set-theoretic representations of models of the theory. Mundy claims that it is by conflating statements in these two languages that one is led into the hole argument, for it is only in the semantic metalanguage that one can express the claim that the ``same'' point has ``different'' metric values in two isomorphic models.  But assertions in the semantic metalanguage do not have physical significance.  They are not assertions within GR.

On its face, Mundy's position is different from those defended in the more recent literature.  Most notably, Mundy's argument apparently turns on the introduction of an intrinsic axiomatization of pseudo-Riemannian geometry; on the other hand, Weatherall's argument is that the hole argument does not go through on the \emph{standard} (modern, coordinate-free) formalism of GR. But in fact, we claim, the arguments are much closer than they appear.

Weatherall argues that, as a principle of mathematical practice, mathematical objects are defined only up to isomorphism, where what counts as an ``isomorphism'' for a given mathematical object is determined by the mathematical theory of those objects. And so, when we aim to represent some physical situation using objects of a certain kind, subject to a given mathematical theory, the ``representational capacities'' of those objects are invariant under isomorphism, meaning that isomorphic objects can represent any given situation equally well.  Inasmuch as the standard of isomorphism for pseudo-Riemannian manifolds is isometry, the models of concern in the hole argument have the same representational capacities. From this he concludes that the hole argument is blocked, because any sense in which the models invoked in the hole argument differ is not reflected in their representational capacities. 

Some parallels with Mundy's argument are immediately apparent.  Like Mundy, Weatherall's key premise is that certain mathematical objects -- viz. pseudo-Riemannian manifolds -- are defined only up to isometry, and for that reason, isometric manifolds are in some sense ``the same''.  Mundy gets to this via an axiomatization and formal semantics; Weatherall gets to it by observing that mathematicians generally intend to attribute to mathematical objects only structure that is preserved by the relevant notion of isomorphism. Both agree that whatever else is the case, putative differences between pseudo-Riemannian manifolds that are not preserved by isometry lack representational, or semantic, significance. Since the hole argument turns on such non-invariant facts, Mundy and Weatherall both conclude that it cannot be stated using just the structure, or language, of GR.

But what about where the hole argument goes wrong?  Here, Weatherall and Mundy appear to differ.  For Weatherall, the problem arises due to a conflation between two different relationships that obtain between $(M,g_{ab})$ and $(M,g_{ab}')$, captured by two different maps between these structures. On the one hand, there is an isometry between $(M, g_{ab})$ and $(M,g_{ab}')$ which is realised by $\psi$. Compared relative to $\psi$, $(M,g_{ab})$ and $(M,g_{ab}')$ are merely two set-theoretic representations of a single pseudo-Riemannian manifold. On the other hand, there is the identity map on $M$, which allows one to talk about particular manifold points. Compared relative to the identity map, $(M, g_{ab})$ and $(M,g_{ab}')$ are not mathematically equivalent. Therefore, although one can say that there are points at which the metric differs when comparing the models using the identity map, one cannot simultaneously maintain that the situations represented by such models are equivalent.  As he puts it, ``There is a sense in which $(M, g_{ab})$ and $(M, g_{ab}')$ are the same, and there is a sense in which they are different. ... But---and this is the central point---one cannot have it both ways'' \citep[p. 338]{WeatherallHoleArg}

This part of Weatherall's argument has been particularly controversial.  After all, in many contexts it is sensible to say that two (distinct) things are the same in some respects and different in others; such relationships can often be expressed via two or more mappings between those things.\footnote{This objection is raised forcefully by \citet{Pooley+Read}.} We think Weatherall's argument may be clarified, here, by reading it as an alternative statement of Mundy's. For Mundy, recall, what goes wrong in the hole argument is that any differences between isometric spacetimes can only be expressed by moving to the semantic metatheory, where one has additional expressive resources. We suggest that this move to the semantic metalanguage corresponds to what Weatherall takes to be required in order to distinguish $(M, g_{ab})$ and $(M,g_{ab}')$ using the identity on $M$; namely, ``one must be invoking some other, additional structure that is not preserved by isometry" (p. 338). This additional structure is precisely the (extensional) structure concerning the domains, $M$, that is referred to by the language of the metatheory of pseudo-Riemannian manifolds.  Meanwhile, the sense in which the models are the same for Weatherall is that they have the same structure as pseudo-Riemannian manifolds. This explains why, in section 5 of his paper, Weatherall argues the hole argument \emph{would} go through if one began with a different kind of mathematical structure, where the points of a manifold were associated with unique labels.

We now turn to the argument from HoTT/UF.  For \citet{Shulman}, the central claim is that HoTT/UF formalizes, using the methods of intensional type theory, the idea that mathematical objects of a given variety -- or, in the formal sense of HoTT, \emph{terms} of a given \emph{type} -- are defined only up to a specified standard of equivalence.  Assertions about those objects within HoTT/UF must be ``covariant'', i.e., preserved, mutatis mutandis, under equivalences between objects of that type.  And for the type ``pseudo-Riemannian manifolds'', the equivalences are precisely isometries.  Thus, as Shulman puts it, ``anything we can say in [HoTT/UF] about [pseudo-Riemannian manifolds] is automatically covariant under isometry'' (p. 53).  It follows that the hole argument \emph{cannot be formulated} within HoTT/UF, because there is no way to express the (non-covariant) claim that the pseudo-Riemannian manifolds $(M,g_{ab})$ and $(M,g_{ab}')$ differ in their assignments of metric values to points.\footnote{\citet{Ladyman+Presnell} argue that in fact, Univalence is not necessary for this conclusion.} Moreover, Shulman argues that the problem with the hole argument is that it illegitimately conflates two different maps from $M$ to $M$, much as Weatherall does (p. 40).\footnote{See also \citet[p. 323]{Ladyman+Presnell}.}  

We take it that this argument is substantially the same as Mundy and Weatherall's. One might respond that there is an important difference: for Shulman, it is HoTT/UF that dissolves the hole argument, whereas neither Weatherall nor Mundy appear to require type theory. But this difference is chimerical. HoTT/UF, for Shulman, is just a way of being precise about the idea that mathematical objects are only defined up to isomorphism---something Mundy and Weatherall defend on other grounds. One way of thinking about the situation is that HoTT/UF provides an alternative to the set-theoretic semantics that Mundy (explicitly) and Weatherall (implicitly) use for representing pseudo-Riemannian manifolds, preventing one from inadvertently shifting into the semantic metalanguage in the way Mundy and Weatherall criticize.

\section{No ``Syntactic'' Solutions?}

In the previous section, we isolated an argument -- the ``mathematical response'' to the hole argument -- offered, in superficially different forms, by several authors.  The mathematical response states that the hole argument, by invoking a privileged identification of spacetime points across models of general relativity, involves assertions, or invokes structure, that go beyond the theory of GR and should be viewed as representationally irrelevant.  

The various articulations of the mathematical response have been criticised on several fronts. Here we consider two such critiques. The first, due to \citet{RynasiewiczSyntax}, is directed at Mundy's version of the argument;\footnote{Rynasiewicz also responds to \citet{Leeds}, who he takes to offer a similarly ``syntactic'' response to the hole argument; in our view, the Leeds response is different and we do not consider it here.} whereas the second, due to \citet{Roberts}, is directed at Weatherall's.  We suggest that both can be rebutted by keeping in mind the mathematical response as just stated.

\cite{RynasiewiczSyntax} attributes to Mundy the view that ``the difficulties allegedly posed [by the hole argument] evaporate if only the theories in question are suitably formulated in an appropriate formal language.  In this sense [he offers] a syntactic solution to the hole argument'' (p. 855).  But the hole argument presents a semantic, or metaphysical, problem: it concerns the relationships between possible worlds.  And one cannot dissolve an interpretational problem through merely ``syntactic'' considerations, or by introducing a novel formalism for expressing the theory. If a given formal language is unable to express an interpretational dilemma, it simply lacks the resources to capture something of manifest, prior significance. Moreover, Rynasiewicz argues, the hole argument \emph{could} be expressed within Mundy's framework if one simply added constants to the theoretical language by which one could specify the coordinates of any point in a given chart.  

A similar charge might be leveled against Shulman's argument, since he, too, invokes a novel formal framework, and then argues that the hole argument cannot be expressed within that framework. But this response seems to mistake the role that formal methods play for Mundy and Shulman.  As we read Mundy, the key issue is not whether the hole argument can be expressed in his preferred axiomatization; rather, it is whether it can be expressed in \emph{any} language for which model isomorphism coincides with isometry.  Shulman, likewise, uses HoTT/UF only to regiment the ideas, already widespread among mathematicians, that mathematical objects are only defined up to some notion of equivalence or isomorphism and reasoning about those objects must be covariant.

Rynasiewicz's suggestion that the theory can be extended, meanwhile, amounts to \textit{changing} GR, so that isometry is no longer the relevant standard of isomorphism.  This sort of response is the same as the one Weatherall contemplates when he argues that in running the hole argument, philosophers implicitly augment the structure of pseudo-Riemannian manifolds to compare points across models. But \citet{Mundy} also anticipates, and rebuts, such a response. He says: ``philosophers are welcome to construct modal extensions of physical theories. However, I claim that nothing in standard physical theory supports such extensions: no scientific problem requires introduction of any primitive relation extending across different models. The fact that transworld identity revives the hole argument is merely another objection to it'' (p. 522). In other words, while one \emph{can} extend the theory to accommodate the hole argument, there are no empirical or scientific justifications for doing so.

Rynasiewicz goes on to observe that Mundy's argument has the following steps.  Suppose that one has a formal semantics for a given theory.  Then, (1) isomorphic models satisfy exactly the same sentences. Hence (2) isomorphic models are semantically equivalent. Therefore (3) isomorphic models represent the same physical situations. Rynasiewicz argues that if Mundy gets from (1) to (2) by defining semantic equivalence as satisfying the same sentences, then (3) does not follow from (2) without further assumptions. More importantly, Rynasiewicz contends that these further assumptions must go past merely formal considerations.

In particular, Rynasiewicz argues that whether two isomorphic models can represent the same situation depends upon whether the isomorphism ``preserves whatever facts must be in place in order to continue to identify the items in the domain of discourse assuming the same repertoire of identification methods" (p. S57), which ``transcends standard formal semantics" (S59).  Rynasiewicz uses the following example to support this claim. Imagine that one has a deck of cards, and a theory that includes predicates for suit, rank, and order in the deck. This theory tells us the standard facts about a deck of cards, such as the distribution of suit and rank, and it asserts that the order is strictly linear.  Now consider a model of this theory, and a map that is a permutation of the domain. One can generate a new model of the theory, isomorphic to the first, via the standard pushforward construction.  Mundy, presumably, would say these two models are semantically equivalent.  But Rynasiewicz argues that whether we should think of these as representing distinct situations depends on the full set of resources one has for identifying objects in the domain of these models, including ones outside the theory such as whether the cards bear additional, observable markers.

We agree that purely formal considerations do not settle the issue.  But Rynasiewicz's example is not analogous to the hole argument. The reason is that Rynasiewicz supposes one can (observationally) distinguish his isomorphic models through features that are not represented in the theory. In other words, the theory is only a partial description of the world. For the theory to be a full description, one must extend the theory to include the further properties under consideration. On the other hand, the hole argument is supposed to present a problem for someone who takes GR to provide a complete description of spatio-temporal structure---that is, a ``realist'' about spacetime in GR.  Such a person is not concerned that the theory cannot express physically significant facts about space and time, much less observable ones; to the contrary, they hypothesize that space and time have just the structure described by the theory of pseudo-Riemannian manifolds, and no more. 

We now turn to \citet{Roberts}, who, echoing Rynasiewicz, writes that whether isomorphic models of a theory represent the same physical situation depends upon ``how we happen to use language to represent the physical world" in a particular domain of discourse.  But his argument is different.  Roberts claims that Weatherall equivocates between two premises.  One, which Roberts accepts as well-motivated, is that isomorphic objects can always represent the same situations.  The second, stronger, view is that isomorphic objects can always represent the same situation ``at once'', or, ``in the same concrete interpretation'', where by this he means relative to a particular specification of the physical entities to which the elements of the domain of the model correspond.

According to Roberts, to block the hole argument one would need to show that any two isometric pseudo-Riemannian manifolds can represent the same physical situation ``at once''.  But this, he argues, cannot be done. Consider the open half-plane $(M, g_{ab})$ where $M=\mathbb{R} \times (0,\infty)$ and $g_{ab}$ is the Minkowski metric restricted to this manifold. Now translate the half-plane up by some amount $a$ into itself to produce the model $(M', g_{ab})$ where $M'= R\times(a, \infty)$. This translated model is isometric to the first. But $(M', g_{ab})$ is also a proper subset of $(M, g_{ab})$, since $(a, \infty)$ is a proper subset of $(0, \infty)$. Since a proper subset of a spacetime represents only ``part'' of the physical situation that the original spacetime does, these two models, despite being isometric, cannot represent the same situation ``at once''. Roberts allows that an applied mathematician can neglect this relationship between the models---but only by allowing the concrete interpretation of the models to vary.  Thus mathematical considerations alone cannot block the hole argument; in addition, one needs to make further interpretational choices about how to use mathematical models in specific representational contexts.  

It should be clear from the foregoing that Roberts' example makes use of the semantic metatheory; one cannot distinguish the open half-plane and its translated model within the theoretical language, since the models are isomorphic.  At issue is whether one \textit{must} invoke the semantic metatheory in the way he does.  Roberts' picture of ``concrete'' representation using a given physical theory seems to be that one first constructs formal models of a theory, and then one identifies elements of the domains of those models with objects in the world.  In other words, for Roberts representation necessarily operates at the level of metatheory, and models of a theory should be taken to have all of the structure describable at the metatheoretic level.  But this is a view of representation that advocates of the mathematical response would reject, on the grounds that the representational capacities of mathematical objects are precisely those preserved by isomorphism.

\section{What Does the Mathematical Response Really Amount To?}

We have argued in the foregoing that Mundy, Weatherall, and Shulman all raise the same objection to the hole argument, despite invoking different mathematical formalisms to do so.  All of them claim that if spacetime has the structure of a pseudo-Riemannian manifold, then, since pseudo-Riemannian manifolds are defined only up to isometry, the differences between isometric models considered in the hole argument have no representational or semantic significance.  We then considered two counterarguments.  Both begged the question against the mathematical response: Rynasiewicz denied that spacetime has the structure of a pseudo-Riemannian manifold and Roberts adopted a theory of representation rejected by Mundy et al.

This discussion suggests that the real issue is what it means to say that some part of the world has a certain structure or is described by some theory.  In this regard, it parallels other, long-standing disputes in philosophy. \citet{RynasiewiczSyntax}, in responding to Mundy, suggests that the hole argument is analogous to the classic problem of reference, raised by \citet{Quine}, \citet{Davidson}, and \citet{Putnam}. Given any model of a formal theory representing some domain of discourse, one can always build an isomorphic model by permuting elements of its domain and then pushing forward the properties, relations, etc. along that permutation.  But, the argument goes, this means the formal theory cannot determine which model assigns these properties, etc. to the ``right'' object.  

For Rynasiewicz, this analogy to the problem of reference shows that whether two models of a formal theory can represent the same situation depends on further, extra-theoretic facts about how we individuate objects.  Roberts does not invoke the problem of reference, but he seems to adopt a similar perspective, since within a given concrete representation, he takes permutations on the domain of a model to generate changes in what physical objects are taken to instantiate what properties.  But the advocate for the mathematical response would presumably deny that there is any problem of reference in the first place, at least for theories that one takes to fully characterize their subject matter.  After all, to generate the problem for any given theory, one must move from the formal theory under consideration to the metatheory.  And the metatheory is representationally irrelevant.  

To deny this in the case of GR is to deny that a century of physics has given us good reasons to believe that space and time have a particular structure, fully characterized by a theory where the correct standard of equivalence for its models is isometry. One chooses such a formalism precisely because there is no physical reason to include more. Anyone who wishes to claim that GR is incomplete in this regard needs to provide some justification for extending the theory.  And it is hard to see why that is motivated.

\section*{Acknowledgments}
This material is partially based upon work produced for the project “New Directions in Philosophy of Cosmology”, funded by the John Templeton Foundation under grant number 61048.  We are grateful to David Mwakima and Jingyi Wu for detailed comments on a previous draft.

\bibliographystyle{elsarticle-harv}
\bibliography{holes2}

\end{document}